\documentstyle[12pt]{JHEP3}
\date{\today} 
 
\def\be{\begin{equation}} 
\def\ee{\end{equation}} 
\def\bear{\begin{eqnarray}} 
\def\eear{\end{eqnarray}} 
\def\nn{\nonumber}

\def\tr{{\mbox{tr}}}

\def\a{\alpha} 
\def\b{\beta} 
\def\g{\gamma} 
\def\u{\mu} 
\def\v{\nu}

\def\Lbd{{\cal L}} 
\def\bj{{\overline{j}}} 
\def\bk{{\overline{k}}} 
\def\bz{{\overline{z}}} 
\def\wdg{{\wedge}} 
\def\wF{{\widetilde{F}}} 
\def\wtau{{\widetilde{\tau}}} 
 
\def\wCF{{\widetilde{\Phi}}} 
\def\wCX{{\widetilde{X}}} 
\def\wCY{{\widetilde{Y}}} 
 
\def\wVF{{\widetilde{V}}} 
\def\th{{\theta}} 
\def\bth{{\overline{\theta}}} 
\def\lam{{\lambda}} 
\def\blam{{\overline{\lambda}}} 
 
\def\cD{{D}} 
\def\bcD{{\overline{D}}} 
\def\cW{{W}} 
\def\bcW{{\overline{W}}} 
 
\def\ad{{\dot{\alpha}}}

\def\rt{{\rightarrow}}

\newcommand\px[1]{{\partial_{#1}}} 
\newcommand\qx[1]{{\partial^{#1}}}

\title{On the Strong Coupling Dynamics of\\
  heterotic string theory on $C^3/Z_3$}

\author{O. J. Ganor\\
\em  Department of Physics\\
University of California, and\\
Theoretical Physics Group \\
Lawrence Berkeley National Laboratory\\
Berkeley, CA 94720\\
Email: \email{origa@socrates.berkeley.edu}}

\author{J. Sonnenschein\\
\em School of Physics and Astronomy\\
Beverly and Raymond Sackler Faculty of Exact Sciences\\
Tel Aviv University, Ramat Aviv, 69978, Israel, and\\
School of Natural Sciences,\\
Institute for Advanced Study\\
Einstein Drive, Princeton, New Jersy, 08540  \\
Email: \email{cobi@ias.edu}}
\abstract{
We study the strong coupling dynamics of the heterotic $E_8\times E_8$ string
theory on the orbifolds $T^6/Z_3$ and  $C^3/Z_3$ using the duality with 
the Horava-Witten  M-theory picture.
 This leads us to a conjecture about the
low energy description of the five dimensional $E_0$-theory (the CFT that describes
the the singularity region of M-theory on $C^3/Z_3$) compactified on $S^1/Z_2$.
}


\received{Feb. 28th, 2002}

\preprint{\hepth{0202206}\\ UCB-PTH-02-07\\ LBNL-49632}
\begin{document} 
\noindent\rule\textwidth{.1pt}\vskip 0.5em
\tableofcontents\vskip 0.5em\noindent\rule\textwidth{.1pt}\vskip 0.5em

\section{Introduction} 
% ========================================================================== 

Recently, models of 5D space-time bounded by two end-of the world branes
attracted  attention both as a laboratory for the 
phenomenology of  elementary  particle physics as well as of 
 novel cosmological scenarios. 
A class of such models is inspired by  
compactifications of the Horava Witten  (HW) M-theory \cite{HW1,HW2}. 
The understanding of the underlying 
5D bulk physics is thus  a key ingredient for the study of such models.

It turns out, however,  that  these 5D bulk theories associated with orbifold compactifications, are generally not well understood. 
This situation is demonstrated by the following puzzle.
Consider the  HW  duals of  the 6d   $T^4/Z_N$ heterotic orbifolds. 
In this case  there is  
 a  localized gauge symmetry $G_1\in E_8$ on  one of the  6D end-of-the-world branes,
and $G_2\in E_8$ on the other one. Generically the heterotic  spectrum 
includes  massless  
twisted particles that seem to be charged under both $G_1$  and $G_2$.      
The puzzle is how to account 
 for these states in the HW picture. 
The resolution of the puzzle follows the realization that in fact there is also
a non-perturbative 7D bulk gauge theory $G_{bulk}$  associated with the 
$A_{N-1}$ singularity of the corresponding ALE space.
The twisted states are in fact charged under say $G_2$ and $G_{bulk}$ \cite{FLO, KSTY}
and not under $G_2$ and $G_1$. 
Proving the consistency of this scenario, namely, that there is a full anomaly cancellation
of local symmetries, requires assigning particular boundary conditions
 to the 6d vector
and hyper multiplets associated with the 7D vector multiplet.
These boundary conditions, which seem to be quite ad hoc
 in the HW picture, turn out to be very  natural when a 
duality  with type I' string is  invoked \cite{GKSTY}. 
In the type I' picture the twisted states can be
traced back to strings associated with brane junctions that involve $D6$ branes,
$D8$ branes $O_8$ orientifold planes and $NS5$ branes.   

 When analyzing the HW duals of 4D heterotic orbifold models, namely, compactifications on
$T^6/Z_N$ one faces a similar puzzle. Again there 
are massless twisted states  that are charged
under  both $G_1$  and $G_2$. 
However, since the geometry at the vicinity of the fixed points 
is now $R^6/Z_N$,
which is not associated with an $A_{N-1}$ singularity but rather
with a strongly coupled $E_0$ theory \cite{Seiberg},
there is no room for a  non-abelian non-perturbative
5D gauge symmetry.
  Thus, the mechanism that resolves the puzzle has to be of a 
different origin.

The goal of this paper is to explore the duality between the heterotic theory on the $T^6/Z_3$
orbifold compactification and M-theory on  
$(S_1/Z_2)\times (T^6/Z_3)$.
In particular we would like to account for the twisted mixed states.

%This implies an analysis of the $E_0$ gauge theory compactified on the interval  ${S_1\over Z_2}$.

The compactification of M-theory on $C^3/Z_3$ was intensively 
explored \cite{Witten:1996qb,Seiberg,MS,KMV,Douglas:1996xp,GMS,IMS}. 
 The corresponding 
low energy field theory is the ``mysterious''
$E_0$ \cite{MS} theory  which is a  strongly coupled 5D CFT with 8 supersymmetries and a 
one dimensional Coulomb branch. 
The $E_0$ theory  has been explored using various  different techniques including
non-trivial fixed points of the renormalization flow of 5D supersymmetric theories
\cite{Seiberg,MS,GMS,IMS}, collapse of de Pezzo surfaces in Calabi-Yau compactifications
\cite{KMV,Douglas:1996xp,MS,CV} and type I' string theories\cite{CV}. In spite of these study efforts 
and due to its strongly coupled nature, the $E_0$ theory is still not well understood.

Even though the full description of the $E_0$ theory is lacking,
 partial results, based on educated guesses,
 about the low energy description of the compactified
theory can be obtained.
This is similar to the situation with the 6D $(2,0)$ theory where
after compactification on $T^2$, 
the low-energy description of the theory is given by the
$N=4$ Super-Yang-Mills field theory.
This field theory is interacting and is believed to correctly describe
all excitations as long as their energy is much lower than the compactification
scale.

In this paper we study the compactification of the $E_0$ theory
on the segment, $S^1/Z_2$, with certain boundary conditions that preserve
$N=1$ supersymmetry in 4D. We will propose a Lagrangian that
(presumably) describes the low energy excitations at a scale below the
compactification scale.

The motivation for this Lagrangian comes from the
 study of the strong coupling dynamics of the heterotic  string
theory on the orbifolds $T^6/Z_3$ and  $C^3/Z_3$.
The notation $C^3/Z_3$ and $T^6/Z_3$ is somewhat ambivalent
because there are several ways to specify the action of $Z_3$
on the $E_8$ gauge degrees of freedom.
In this paper we concentrate mainly on the orbifolds 
that break the $E_8\times E_8$ gauge group down to
$SU(3)\times E_6\times SU(3)\times E_6$.
We assume that the volume of $T^6/Z_3$ is large so that worldsheet
instantons can be neglected.
We analyze the moduli space of these orbifolds from the heterotic string 
and the low energy supergravity pictures. 
We discuss the local anomaly cancellation in the various scenarios.

The paper is organized as follows.  The moduli space
of the $T^6/Z_3$ orbifold  is discussed  in section 2 from the heterotic theory point of view.
We start with a brief description of the model, its spectrum, superpotential
and D- term. We then analyze the F-term flatness condition for the
$R^6/ Z_3$ case, and this condition combined with the D-term flatness for the compact case. We show that the moduli space for the non-compact case is
a blow-down at the zero section of a certain line bundle over
$P^2\times P^2$.
Section 3 is devoted to a brief reminder of the geometry of the blow-up of the fixed point
of the $C^3 / Z_3$ orbifold. 
In particular the metric, complex structure and the Euler number are written down.
The moduli space is then reproduced from the supergravity description in the large blow-up limit.
For completeness, we discuss gauge instantons for our model as well as the other  $T^6 / Z_3$ 
orbifolds.
The strong coupling limit as inferred from the Horava Witten dual theory is the topic of section 6.
We identify the two scales in the systems, namely, the compactification scale and the scale of the
expectation value of the scalar field.
We write down the
 $N=1$ supersymmetric 5d  $E_0$  theory  in terms of a 4D $N=1$ chiral and vector superfields.
We then compactify this theory on $S^1/Z_2$ first in the limit of an expectation value which is   much larger  than the inverse of the compactification scale.  In this regime we reduce the 
11D HW supergravity to that of a 5D theory  in the form of a non-linear sigma model. We then rewrite
it in terms of a linear sigma model and determine the relations between the linear an non-linear
descriptions.  We then conjecture about the theory in the opposite regime where the compactification scale is larger than the inverse of the expectation value of the scalar field. 
 Section 6 is devoted to a discussion of the anomaly cancellation in  both the compact and non
compact cases. In the former case the cancellation is between the contribution of the 
twisted states and that of the untwisted states after division  
 by the number of fixed points. In the latter case the integration over  the zero mode of the orbifold 
operation results in an identical division. 
In section 7 we summarize our results and state several open questions.       
\section{The moduli space from heterotic string theory} 
% ========================================================================== 
\subsection{The model}\label{subsec:model} 
% -------------------------------------------------------------------------
The model is the  heterotic string  on a $T^6/Z_3$ orbifold.
The $T^6$ is of the form $T^2\times T^2\times T^2$ where each 
$T^2$ is a quotient of the complex plane with the root lattice  of $SU(3)$, namely, 
$T^2={C\over \Lambda_{SU(3)}}$.
The tori are 
  characterized by the 
 identifications $z_i\sim z_i+1$ and $z_i\sim z_i+ e^{\pi i/3}$ where $i=1,2,3$ and
 admit a $Z_3$ generated by  the transformations
\be
\Omega_1:\  z_i\rt \alpha(z_i)e^{2\pi r_i/3}z_i;\qquad r_i = (1,1,-2)   
\ee
 There are 27 fixed points of  the $\alpha_i$
 at $z_i=0, z_i= 1/\sqrt{3}e^{\pi i/6},
z_i= 2/\sqrt{3}e^{\pi i/6}$. 

In addition to the vector $r_i$ the orbifolding operation is characterized also by
its shift vector $s_K$, $k=1,...16$ 
defined by the transformation of the 16 complex left-moving fermions 
\be\label{omega2}
\Omega_2:\ \lambda^{K\pm}\rightarrow e^{\pm {2\pi i s_K\over 3}}\lambda^{K\pm}
\ee
Then we define $\Omega = \Omega_1\circ \Omega_2$ to be the generator 
of the $Z_3$ orbifold.
The shift vector associated with the current model  
$s_K=(1,1-2,0^5;1,1,-2, 0^5)$, implies the breaking of 
each $E_8$ factor down to $E_6\times SU(3)$ 
(for more details see appendix A).

\subsection{The spectrum, superpotential and D-term} \label{subsec:spectrum}
% -------------------------------------------------------------------------
The spectrum of the model contains the untwisted states
\be\label{untwisted}
3(3,27,1,1)\oplus 3(1,1,3,27)\oplus 9\ moduli
\ee
and the twisted states
\be\label{twisted}
27(\overline{3},1,\overline{3},1)
\ee
where the decomposition is under $SU(3)\times E_6\times SU(3)\times E_6$.
(In Appendix A the spectra of the other  $T^6/Z_3$  models  is presented).

Let $A=1\dots 27$ be a label of the fixed-point of $Z_3$ inside $T_6/Z_3$. 
Near such a fixed point the space looks like $R^6/Z_3$. 
We have one chiral field, $\Phi_A$, in the $(\overline{3},1,\overline{3},1)$ 
localized around each fixed point. 
The moduli space is determined by the F-term and D-term conditions. 
The F-term comes from the superpotential: 
$$ 
W(\Phi_1,\dots,\Phi_{27}) = \sum_A \det\Phi_A, 
$$ 
where we think of $\Phi_A$ as a $3\times 3$ matrix. 
The D-term conditions are: 
$$ 
g_{\tiny YM}^2 \tr\sum_A \Phi_A^\dagger \tau^a\Phi_A = 0,\qquad 
g_{\tiny YM}^2 \tr\sum_A \Phi_A \tau^a\Phi_A^\dagger = 0,\qquad a=1\dots 8, 
$$ 
where $\tau^a$ is a generator of $SU(3)$ (taken as a $3\times 3$ matrix) 
and $g_{YM}^2$ is the 4D $E_8$ coupling constant. 
It is given by the 10D coupling constant divided by the volume of 
$T^6/Z_3$.

\subsection{The F-term for the orbifold $R^6/Z_3$}\label{subsec:Fterm}
% -------------------------------------------------------------------------
Most of  the nontrivial dynamics is localized at the fixed points. 
So we analyze $R^6/Z_3$ first. 
Since the volume of $R^6$ is infinite we can set $g_{YM} = 0$ and 
forget about the D-terms. 
There is only one $3\times 3$ field $\Phi$ with superpotential 
$W\equiv \det\Phi$. 
The F-term constraints are $W' = 0$ where $W'$ is the matrix  
of $2\times 2$ minors of $\Phi$. 
Thus, the F-term constraints imply that the rank of $\Phi$ is at most 
$1$. 
We can therefore write $\Phi$ as: 
\be\label{uvdag} 
\Phi = u v^T, 
\ee
where $u$ is a vector in the $(\overline{3},1)$ of $SU(3)\times SU(3)$ 
and $v$ is a vector in the $(1,\overline{3})$. 
$u$ breaks the left $SU(3)$ down to $SU(2)$ and $v$ does 
the same to the right $SU(3)$ so altogether we are left with 
$SU(2)\times SU(2)\times U(1)$ where $U(1)$ acts as: 
\be\label{uvch}
u\rightarrow e^{i\theta} u,\qquad 
v\rightarrow e^{-i\theta}v. 
\ee

\subsection{The F-term and D-term for the orbifold $T^6/Z_3$} 
% -------------------------------------------------------------------------
Moving back to $T^6/Z_3$, we solve the individual F-term constraints and 
find that each $\Phi$ can be written as 
$\Phi_A = u_A v_A^T$ ($A=1\dots 27$). 
The D-term constraints imply 
$$ 
0 = \sum_A (u_A^\dagger u_A)(v_A^\dagger \tau^a v_A) 
 = \sum_A (v_A^\dagger v_A)(u_A^\dagger \tau^a u_A),\qquad a=1\dots 8 
$$ 
This can only be satisfied if 
$$ 
\sum_A (v_A^\dagger v_A)u_A u_A^\dagger = 
\sum_A (u_A^\dagger u_A)v_A v_A^\dagger = c I. 
$$ 
In particular, unless all of the $u_A$'s and $v_A$'s are zero, 
we need at least 3 different $A$'s with nonzero $u_A$ and $v_A$. 
This is because the matrix $u_A u_A^\dagger$ is of rank $1$ and 
the sum of 2 matrices of rank $1$ can never be $c I$ (i.e. of rank $0$) 
for $c\neq 0$. 
 
Note that $(\Lambda_1,\Lambda_2)\in SU(3)\times SU(3)$ act as 
$$ 
u_A\rightarrow \Lambda_1^\dagger u_A,\qquad 
v_A\rightarrow \Lambda_2 v_A. 
$$ 
If exactly 3 $A$'s have nonzero $\Phi_A$'s (say $A=1,2,3$) 
then we can use the   
$SU(3)\times SU(3)$ gauge freedom to turn 
the $u_A$'s and $v_A$ into the following form: 
$$ 
u_1 = v_1 = \left(\begin{array}{c} 1 \\ 0 \\ 0 \\ \end{array}\right),\qquad 
u_2 = v_2 = \left(\begin{array}{c} 0 \\ 1 \\ 0 \\ \end{array}\right),\qquad 
u_3 = v_3 = \left(\begin{array}{c} 0 \\ 0 \\ 1 \\ \end{array}\right). 
$$ 
In this case $\Phi_1,\Phi_2,\Phi_3$ are all diagonal and 
the unbroken symmetry is $U(1)^2$. 
If $l>3$ in general, all of the $SU(3)^2$ will be broken.

\subsection{The moduli space for $R^6/Z_3$}\label{subsec:modsp} 
% -------------------------------------------------------------------------
We have seen that the moduli space for the case of $R^6/Z_3$ 
is the moduli space of all $3\times 3$ matrices of rank $\le 1$. 
we can parameterize it as follows. 
Take $\Phi = u v^T$ as before.  
$u$ and $v$ are not uniquely defined because we can 
change $u\rightarrow\lambda u$ and at the same time 
$v\rightarrow \lambda^{-1} v$. 
If $\Phi\neq 0$ then neither 
$u$ nor $v$ are zero. The equivalence relation $u\sim \lambda u$ (for 
$\lambda\in C$) defines a point on $P^2$ that we will denote by  
$[u]$. Similarly $v$ defines a point $[v]$ in $P^2$. 
For fixed $[u]$ and $[v]$ 
$\Phi$ has one more complex degree of freedom which is the overall 
scale and we denote it by 
$e^{\sigma}$ (for $\sigma\in C$). 
As $\sigma$ varies,  
$e^{\sigma}$ spans a plane (complex line) 
which is fibered over $P^2\times P^2$. 
Let $\Lbd_1$ ($\Lbd_2$) be the universal line bundle over 
the first (second) $P^2$. 
Then, 

{\it  \centerline{ The moduli space is the blow-down of} 
\centerline{$\Lbd_1\otimes\Lbd_2$ over $P^2\times P^2$ at the zero section.}} 
 
We wish to understand how this moduli space can be interpreted in 
the limit of small curvature and small gauge field strengths. 
In this case the moduli space is understood as the moduli space of 
the deformation of the geometry into a nonsingular space and a nonsingular 
instanton configuration. 
The space $R^6/Z_3$ can be deformed into a smooth space with 
the singularity at the origin replaced by $P^2$. 
Our goal is to argue that $e^\sigma$ parameterizes the size of the  
$P^2$ (and the $B$-field on it) and $[u]$ and $[v]$ parameterize 
the $E_8\times E_8$ instanton configuration.

\section{Some facts from geometry} 
% ========================================================================= 
We need some facts about the blow-up of $C^3/Z_3$ and 
the normalizable harmonic 2-form on it. 
In general $C^n/Z_n$ (where the generator of 
$Z_n$ acts as multiplication by 
$e^{\frac{2\pi i}{n}}$ on all the coordinates of $C^n$) can be deformed  
to a smooth (noncompact) Calabi-Yau manifold that approaches 
$C^n/Z_n$ at infinity. In this CY the singular point at the  
origin on $C^n/Z_n$ is replaced with a compact projective space 
$P^{n-1}$. 
 
\subsection{Metric and complex structure} 
% -------------------------------------------------------------------------
We can take the metric on $P^{n-1}$ to be the Fubini-Study metric 
$$ 
g_{i\bj} = \frac{\delta_{i\bj}(1+\sum|z_k|^2)-z_i\bz_\bj}{ 
  \left(1+\sum|z_k|^2\right)^2}, 
$$ 
which is derived from the K\"ahler function 
$$ 
K = \log\left(1 +|z_1|^2 + \cdots +|z_{n-1}|^2\right). 
$$ 
The normalized $(1,1)$ form is $\omega = i g_{i\bj}dz_i\wdg d\bz_\bj$. 
It can be written as $\omega = dA$ with 
$$ 
A = \frac{i}{2}e^{-K}\sum_k (z_k d\bz_\bk - \bz_\bk dz_k). 
$$ 
The blowup of $C^n/Z_n$ can be written as a line bundle 
over $P^{n-1}$ with first Chern class $c_1 = -n\omega$. 
Choose a constant $a>0$. 
Let $z_1,\dots,z_{n-1}$ parameterize the point on the base 
and let $0\le\theta<2\pi$ be a periodic variable and 
$a\le r <\infty$ be a real coordinate. 
The metric on the blown-up space can be written as 
\bear 
ds^2 &=&  
 \frac{1}{2}r^2 \sum_{i,\bj=1}^{n-1} g_{i\bj} dz_i d\bz_\bj 
\nn\\ &+& 
\left(1 - \frac{a^{2n}}{r^{2n}}\right)^{-1} dr^2  
+\frac{r^2}{n^2}\left(1 - \frac{a^{2n}}{r^{2n}}\right)(d\theta - n A)^2 
\nn 
\eear 
The holomorphic coordinates are $z_1,\dots,z_{n-1}$ and 
$$ 
 w = \sqrt{r^{2n}-a^{2n}} e^{\frac{K}{2} + i \theta}. 
$$ 
This space has a normalizable 2-form: 
\be\label{twof} 
\wF = \frac{2(n-1)a^{2(n-1)}}{n r^{2n-3}} dr\wdg (d\theta -n A) 
 +\frac{a^{2(n-1)}}{r^{2(n-1)}}\omega. 
\ee 
It satisfies 
\be\label{holof} 
\int_{P^1\subset P^{n-1}} \wF = 2\pi,\qquad 
\int_{{\mbox{Fiber}}} \wF = \frac{2\pi}{n}. 
\ee 
Also 
\be\label{fff} 
\int\wF\wdg\wF\wdg\cdots\wdg\wF = \frac{2\pi}{n^2}\int\omega^{n-1} 
=\frac{(2\pi)^n}{n^2} 
\ee 
Here $P^{n-1}$ denotes the divisor given by $r=a$ and 
$P^1\subset P^{n-1}$ is any (complex) line inside this divisor. 
The fibers are given by fixed $z_k$'s. 
For $n=3$ we can calculate 
$$ 
-\frac{1}{2}\int_{P^2}\tr_6 R\wdg R = 12 (2\pi)^2. 
$$ 
The trace is in the representation $6$ of $SO(6)$. 
 
Another property of interest is the Euler number of the $T^6\over Z_n$ 
orbifolds\cite{DHVW}.
Consider in particular the $n=3$ case.
At each  fixed point  one deletes the singular point and glues instead, as was explained above,
  a non-compact 
manifold composed of  a  $P^2$ and 
  a line bundle over it  with a resulting  $c_1=0$.
The Euler number of each of these glued manifolds is $\chi=3$. Since $T^2$ has a vanishing $\chi$, the torus with the 27 deleted points has $\chi=-27$. Thus, the Euler number of 
${T^6\over Z_3}$ is  $\chi =-27/3=-9$. We add now the Euler number of the glued manifolds and we end with $\chi =-9+27\times 3 = 72$.
 
\section{The moduli space from supergravity}\label{ModSugra} 
% ========================================================================= 
We would like to reproduce the moduli space found in section 
(\ref{subsec:modsp}) in the limit that the blow-up parameter, 
$a$, is large compared to $l_s$. The moduli space is then 
the moduli space of $E_8\times E_8$ instantons  on the blown-up 
$C^3/Z_3$. The requirement on the instanton is that the holonomy 
of the gauge field at infinity is known.  
Namely, define the contour 
$$ 
\gamma(t)=(z_1=\xi_1,\dots,z_{n-1}=\xi_{n-1},\theta=t, 
 r=r_0),\qquad 0\le t<2\pi. 
$$ 
where $\xi_1,\dots,\xi_{n-1},r_0$ are constants. 
We require that as $r_0\rightarrow\infty$ the holonomy of 
the $E_8\times E_8$ gauge field along $\gamma(t)$ is conjugate to 
$\Omega_2$ that was defined in (\ref{omega2}). 
The second requirement is that  
$$ 
\frac{1}{30}\int_{P^2} \sum\tr_{248} F\wdg F =  
\int_{P^2} \tr R\wdg R = -24 (2\pi)^2. 
$$ 
Here $P^2$ is the exceptional divisor of the  blow-up. 
The instanton solutions that we will consider are abelian. 
They are given by picking a generator $\tau$ of the Lie algebra 
of $E_8\times E_8$, normalized such that $e^{2\pi i\tau}$ is 
the unity in the group, and setting the $E_8\times E_8$ field-strength 
to be $F = i\wF\tau$, where $\wF$ is the 2-form found in (\ref{twof}). 
(Note that $F$ is defined to be anti-hermitian.) 
According to (\ref{holof}), the holonomy around $\gamma(t)$ is 
$e^{\frac{2\pi i}{3}\tau}$. Thus $\tau$ should be chosen so that 
$e^{\frac{2\pi i}{3}\tau}$ will be $\Omega_2$ -- the generator of $Z_3$ 
in the gauge group. The instanton condition implies that 
$$ 
24 = \frac{1}{30}\sum\tr_{248} \tau^2. 
$$ 
For the $\Omega_2$ that breaks $E_8\times E_8$ to 
$(E_6\times SU(3))^2$ we can pick $\tau$ as follows.
Let $\phi_1:SU(3)\rightarrow E_8$ be the embedding of $SU(3)$ in 
the first $E_8$ factor and let $\phi_2$ be the embedding in the second 
factor. 
We pick $\wtau\in SU(3)$ to be ${\mbox{diag}}(1,1,-2)$ 
and take $\tau=\phi_1(\wtau)\oplus\phi_2(\wtau)$. 
Note that for elements of $E_8$ that are embedded in $SU(3)$ 
we have  
$$ 
\frac{1}{30}\tr_{248}\phi_i(\wtau)^2 = 
2\tr_{3}\wtau^2 =12. 
$$ 
Summing the contributions of the two $E_8$ factors gives $24$. 
The moduli space is the moduli space of embeddings 
of $\tau$ inside $E_8\times E_8$ keeping 
the holonomy at infinity, $e^{\frac{2\pi i}{3}\tau}$ fixed. 
The moduli space of different $\wtau$'s in $SU(3)$ 
that are conjugate to ${\mbox{diag}}(1,1,-2)$ is $P^2$. 
Thus the moduli space of instantons is $P^2\times P^2$. 
The coordinate $\sigma$ from (\ref{subsec:modsp}) is interpreted 
as $\sigma = \frac{1}{2}\a'\pi a^2 + i B$ where $\pi a^2$ is the area 
of a $P^1$ divisor inside the exceptional $P^2$ and 
$B$ is the integral of the NSNS 2-form on $P^1$. 
 
It is interesting to check other embeddings of $Z_3$ inside 
$E_8\times E_8$. 
In one embedding we take the generator of $Z_3$ to 
be in the center of $SU(9)/Z_3\subset E_8$. 
The element is $e^{\frac{2\pi i}{9}}$. 
We then take $\wtau\in su(9)$ to be  
$$ 
\wtau={\mbox{diag}}\left( 
\underbrace{\frac{1}{3},\dots,\frac{1}{3},}_{8} 
-\frac{8}{3}\right). 
$$ 
We let $\phi:su(9)\rightarrow E_8$ be the embedding and we calculate 
$$ 
\frac{1}{30}\tr_{248} \phi(\wtau)^2 
=2\tr_{9}\wtau^2 =16. 
$$ 
Therefore, in the second $E_8$ factor we should find an 
embedding with instanton number $8$. 
For this we break $E_8$ to $SU(14)\times U(1)$ and 
embed $\tau$ as $e^{\frac{2\pi i}{3}}$ inside $U(1)$. 

\section{Strong coupling limit}
% =========================================================================
At strong coupling, the $E_8\times E_8$ heterotic string theory
on $C^3/Z_3$ is described by weakly coupled 11D gravity on the bulk of
$(S^1/Z_2)\times (C^3/Z_3)\times R^{4,1}$ and $G\in E_8$ gauge
fields on the two boundaries. The only strongly coupled part of
this background comes from the fixed point of the $Z_3$ action.
It is described by the $E_0$ theory compactified on $S^1/Z_2$.

The $E_0$ theory is a strongly coupled 5D CFT\cite{MS} that describes the
localized degrees of freedom of M-theory on $C^3/Z_3$ at the fixed point.
We will recall some known facts about the $E_0$-theory in
subsection (\ref{ezfacts}).
We will then use the heterotic string analysis of the orbifold
to make conjectures about the 4D low-energy description of
the $E_0$ theory compactified on $S^1/Z_2$.

\subsection{The setting}
% ------------------------------------------------------------------------
The $E_0$-theory is compactified
on an interval $S^1/Z_2$ and, as we shall see,
 there are extra degrees  of freedom on the boundaries.
Let the length of the interval be $\pi R$.
The 5D $E_0$-theory has a moduli space $R/Z_2$ \cite{Seiberg}
that is parameterized by an order parameter $\chi_0$ with dimensions
of mass (see subsection (\ref{ezfacts}) below).

Our problem has two scales: 
\begin{itemize}
\item
The {\it compactification scale},
$\frac{1}{R}$, 
\item 
The {\it $E_0$-scale}, $\chi_0\equiv \langle\chi\rangle$.
\end{itemize}

If $\chi_0 \gg \frac{1}{R}$
the 5D compactification (energy) scale is low and we can first reduce the
$E_0$-theory to its 5D low-energy description and then compactify
the latter on $S^1/Z_2$ with appropriate boundary conditions.

If, on the other hand, the condition 
$\chi_0\gg \frac{1}{R}$ is not met, quantum corrections are
strong and we do not know the metric on the moduli space of the
effective 4D low-energy theory.
Since we do not know any microscopic definition for $\chi_0$
it does not make sense to use $\chi_0$ as a coordinate
on the 4D moduli space.
We can still look for
an effective 4D low-energy description but it requires a better
understanding of the $E_0$ theory.
We will propose a conjecture about that description in section (\ref{conjecture}).

\subsection{The $E_0$-theory}\label{ezfacts}
% -------------------------------------------------------------------------
The $E_0$-theory is a five-dimensional interacting CFT.
It has $N=1$ supersymmetry (8 generators) in 5D.
It has a one-dimensional Coulomb branch parameterized by a real
coordinate $\chi_0 > 0$. For a generic $\chi_0$ the low-energy description
is a 5D vector multiplet with (five dimensional) $N=1$ supersymmetry.
This multiplet comprises of a scalar $\chi$ (whose VEV is $\chi_0$),
a vector field $A$ and fermions.
The low-energy effective action is
\be\label{ezle}
L = \frac{1}{4\pi^2}\int \left\lbrack
\frac{1}{2}\chi\px{\u}\chi\qx{\u}\chi +\frac{1}{4}\chi F_{\u\v} F^{\u\v}
 +\frac{1}{24}
  \epsilon^{\u_1\u_2\u_3\u_4\u_5} A_{\u_1} F_{\u_1\u_2} F_{\u_3\u_4}
\right\rbrack d^5 x
+{\mbox{(fermions).}}
\ee
Here $\chi$ is a scalar field of dimension $1$, $A$ is a low-energy
$U(1)$ gauge field and $F=dA$.
The Coulomb branch is parameterized by the 
VEV $\chi_0 = \langle\chi\rangle$.
The $E_0$-theory has a rich structure of electric BPS particles 
with masses that are integer multiples of $\chi_0$ and magnetic
BPS strings with tensions that are integer multiples of $\chi_0^2$
\cite{Witten:1996qb,KMV,Seiberg}.

We can rewrite
(\ref{ezle}) using 4D $N=1$ superspace.
We separate one direction out of the five, call it $x_4$, and
we define the chiral superfield $\wCF(x_4)$ and the vector
superfield $\wVF(x_4)$ (which depend explicitly
on the parameter $x_4$) so that
\be\label{wCFdef}
\wCF(x_4)|_{\th=\bth=0}= \chi(x_4) + i A_4(x_4)
\ee
and, in the WZ gauge,
\be\label{wVFdef}
\wVF = -\th\sigma^\u\bth A_\u 
  +i\th^2\bth\lam -i\bth^2\th\blam +\frac{1}{2}\th^2\bth^2 D.
\ee
The gauge freedom is
$$
\wVF \rightarrow \wVF + \Lambda + \Lambda^\dagger,\qquad
\wCF\rightarrow \wCF + 2\px{4}\Lambda.
$$
where $\Lambda$ is a superfield.
Without any boundary conditions the Lagrangian is
\bear
L &=&
-\frac{1}{12\pi^2}\int 
\left\lbrack
\frac{1}{8}(\wCF +\wCF^\dagger -2\px{4}\wVF)^3  
+\wVF\px{4}\wVF\cD_\a\cW^\a 
+\px{4}\wVF\bcD^\ad\wVF\bcW_\ad 
+\px{4}\wVF\cD_\a\wVF\cW^\a
\right\rbrack
d^4\th
\nn\\ &&
-\frac{1}{16\pi^2}\int \wCF \cW_\a \cW^\a d^2\th
-\frac{1}{16\pi^2}\int \wCF^\dagger \bcW^\ad \bcW_\ad d^2\bth
\nn
\eear
where, as usual, $\cW_\a = -\frac{1}{4}\bcD^2\cD_\a\wVF$
and it satisfies $\cD^\a \cW_\a = \bcD_\ad\bcW^\ad$.

It is not hard to check that this Lagrangian is gauge invariant up
to a total derivative
\bear
\delta \int L d^4 x &=&
-\frac{1}{24\pi^2}\px{4}\int\Lambda\cW_\a\cW^\a d^2\th d^4 x
-\frac{1}{24\pi^2}\px{4}\int\Lambda^\dagger\bcW^\ad\bcW_\ad d^2\bth d^4 x
\nn
\eear
Expanded in components we find
\bear
L &=& \frac{1}{4\pi^2}\int \left\lbrack
\frac{1}{2}\chi(\px{\u}\chi)^2 
 + \frac{1}{2}\chi F_{4\u} F^{4\u}
+\frac{1}{2}\chi^2\px{4}D -\frac{1}{2}\chi  F^* F
\right\rbrack d^5 x
\nn\\
&+&\frac{1}{4\pi^2}\int \left\lbrack
-\frac{1}{2}\chi D^2
+ \frac{1}{4}\chi F_{\u\v} F^{\u\v}
+\frac{1}{8}A_4 \epsilon^{\u\v\rho\tau}F_{\u\v} F_{\rho\tau}
-\frac{1}{6} A_\u\epsilon^{\u\v\rho\tau}\px{4}A_\v F_{\rho\tau}
\right\rbrack d^5 x
\nn\\
&& \label{fivedl}
\eear
Here $\u=0\dots 3$.
Although this action does not look manifestly Lorentz invariant,
if we eliminate the auxiliary fields we find $D = -\px{4}\chi$ and $F=0$
and the whole action can be rewritten as (\ref{ezle})
and the action is manifestly Lorentz invariant (although supersymmetry
is not manifest).

So far we have discussed the low-energy effective action
in 5D. Later on we will discuss a compactification
on $S^1/Z_2$ that breaks half the supersymmetry.
For completeness, we note that the simplest compactification
of the $E_0$ down to 4D is on $S^1$ in such a way that preserves
$N=2$ supersymmetry. We will take the coordinate along $S^1$
to be $0\le x_4\le 2\pi R$.
The low-energy description of the $E_0$-theory compactified on 
$S^1$ is given by a single $N=2$ $U(1)$ vector multiplet 
with the effective
action that is derived from the Seiberg-Witten curve:
$$
y^2 = (x^2-u)(x-u^2).
$$
Here $u$ is a holomorphic parameter on the moduli space.
For $|u|\rightarrow\infty$ we can write
$$
\frac{1}{2\pi R}\log u  \approx \langle\chi + i A_4\rangle.
$$

\subsection{Adding a boundary}
% ------------------------------------------------------------------------
We take the effective action (\ref{ezle}) and compactify it 
on the interval $S^1/Z_2$. Let $0\le x_4\le \pi R$ be the coordinate
along the interval (the other coordinates are $x_0,\dots,x_3$).
Now we add the boundaries at $x_4 = 0$ and $x_4=\pi R$.
Our goal is to propose a Lagrangian that describes the theory
at energies 
$$
E\ll \frac{1}{R}\ll\chi_0 \equiv \langle\chi\rangle.
$$
In the context of M-theory, as we explained above, the compactification
arises from M-theory on a blown up $C^3/Z_3$ where the singularity
was blown up to a $P^2$ with volume $M_p^{-6}\chi_0^2$.
We are interested in the regime $\chi_0\ll M_p$ where the scale of
our low-energy $E_0$-theory is lower than $M_p$.

The $U(1)$ gauge group of the low-energy effective action (\ref{ezle})
has an anomaly when we add a boundary.
This anomaly comes from the Chern-Simons term 
\be\label{CSterm}
\frac{1}{32\pi^2}\int A\wdg F\wdg F 
\equiv
\frac{1}{32\pi^2}\int \epsilon^{\a\b\g\u\v} A_\a F_{\b\g}F_{\u\v} d^5 x.
\ee
This term comes from the Chern-Simons term of 11D supergravity
reduced on the blown-up $C^3/Z_3$ with
the substitution $C = 3 A(x_0,\dots x_4)\wdg \wF(x_5\dots x_{10})$.
Here $\wF$ is the harmonic 2-form defined in (\ref{twof}).
The factor of 3 is needed because it is $3\wF$ that has integral periods
according to (\ref{fff}).
In 5D, the Chern-Simons term must be an integer product of
$\frac{1}{96\pi^2}\int A\wdg F\wdg F$ (see \cite{Witten:1996qb}).
Our model has 3 times the fundamental unit of anomaly.

There are two possible ways to eliminate the anomaly.
We can either add degrees of freedom on the boundary with 
the opposite anomaly or we can set the boundary conditions
on the gauge field so that its parallel components vanish on
the boundary. In that case, the gauge group contains only transformations
that vanish on the boundary and there is no anomaly.

In the next subsections we will discuss two descriptions of
the theory with a boundary.
We will start with an analysis of the regime 
$\frac{1}{R}\ll M_p\ll\chi_0$ where the Horava-Witten
supergravity approximation to M-theory is valid.
We will see that the dimensional reduction of the Horava-Witten 
low-energy effective action leads to a description where
the gauge transformation vanish on the boundary.
Extra degrees of freedom on the boundary are then described
by a nonlinear $\sigma$-model.

We will then discuss an alternative description in terms of
a linear $\sigma$-model where the gauge transformations are
allowed not to vanish on the boundary. Instead extra charged
chiral superfields cancel the anomaly on the boundary.

\subsection{Description in terms of the nonlinear $\sigma$-model}
% ------------------------------------------------------------------------
When $M_p\ll\chi_0$ the volume of the $P^2$ blow-up is large
and we can reduce the Horava-Witten supergravity Lagrangian --
the 11D supergravity in the bulk and $E_8$ gauge fields on the 10D
boundary -- down to the noncompact 5D.
The reduction of the $E_8$ gauge fields
was discussed in detail in section (\ref{ModSugra}) and the result was
that the low-energy degrees of freedom describe the moduli space
of a certain instanton. The moduli space was described as the embedding
of $U(1)\in SU(3)$ where $SU(3)$ was a fixed subgroup of $E_8$.
The $E_8$ instanton moduli space in this problem is therefore a copy
of $P^2$ (not to be confused with the blow up divisor which is also
a $P^2$ but in space-time and not in field space).
Thus, at low-energies, the degrees of freedom on the boundary are
described by a nonlinear $\sigma$-model with $P^2$ as the
target space. The appropriate boundary conditions for the superfields
$\wCF$ and $\wVF$, defined in (\ref{wCFdef}-\ref{wVFdef}),
can be obtained from
the Horava-Witten boundary conditions on the 3-form field $C$ of 11D
supergravity. Recall that Horava and Witten described the segment
$[0,\pi R]$ as $S^1/Z_2$ with the $Z_2$ acting as
\bear
Z_2: C_{10\u\v}(x_{10}) &\rightarrow& C_{10\u\v}(-x_{10}),\nn\\
Z_2: C_{\sigma\u\v}(x_{10}) &\rightarrow& -C_{\sigma\u\v}(-x_{10}),\nn
\eear
Here $\u,\v=0\dots 9$. In our problem $x_{10}$ should be replaced
with $x_4$. The connection between $C$ and the components 
$A_\u$ and $A_4$ of $\wVF$ and $\wCF$ is
$$
C_{\u IJ} = A_\u \wF_{IJ},\qquad
C_{4 IJ} = A_4 \wF_{IJ}.
$$
where $\wF$ is the 2-form on the blown up $C^3/Z_3$ that was defined 
in (\ref{twof}).
It follows that $Z_2$ acts on $A_\u$ and $A_4$
as 
$$
A_{\u}(x_4) \rightarrow -A_{\u}(-x_4),\qquad
A_4(x_4) \rightarrow A_4(-x_4).
$$
These rules can be extended to the superfields and
we can now derive the boundary conditions
$$
\wVF|_{x_4 = 0} = \wVF|_{x_4 = \pi R} = 0,\qquad
\px{4}\wCF|_{x_4 = 0} = \px{4}\wCF|_{x_4 = \pi R} = 0.
$$

\subsection{The linear $\sigma$-model}\label{LinSM}
% ------------------------------------------------------------------------
We will now discuss an alternative description of the boundary.
The Chern-Simons term (\ref{CSterm}) produces an anomaly
for the $U(1)$ gauge transformations on the boundary.
This anomaly can be canceled by 3 4D chiral fermions with 
$U(1)$ charge $1$ on the boundary.
 Thus, we assume that on the boundary there is
a chiral superfield $\wCX$ that is in the fundamental representation 
$3$ of a global $SU(3)$ and is also charged under the 5D $U(1)$ gauge
field.
The total action now has an extra term
\be\label{Xbound}
\int \wCX^\dagger e^\wVF\wCX d^4\th d^4 x
\ee
On the other hand we do not impose Dirichlet boundary conditions
on the gauge fields.
When we integrate out the auxiliary field $D$ from (\ref{fivedl})
we need to integrate by parts over $dx_4$. When there is a boundary
we find a boundary contribution of
$$
\frac{1}{2}\chi^2 D.
$$
The field $D$ also appears linearly in (\ref{Xbound}). Integrating $D$ out
we find the boundary condition
\be\label{Xbc}
\frac{1}{2}\chi^2|_{x_4 = 0} = |X|^2,\qquad
X\equiv \wCX|_{\th=\bth=0}
\ee
where $X$ is the scalar ($\th=\bth=0$) component of $\wCX$.
Note also that the a single $U(1)$ 
charged chiral superfield is anomalous in 4D but the anomaly is canceled
by the inflow from the term $\int A\wdg F\wdg F$ in (\ref{fivedl}) in 5D.
This consideration also shows that we need exactly 3 chiral fields
to cancel the anomaly from the 5D WZ term.

In the regime $\chi_0\gg\frac{1}{R}$ the other boundary is far and
the effective action near the boundary cannot have any dimensionful
parameter in it.
The superfields $\wCF$ and $\wCX$ both have mass dimensions of $1$.
This implies that there is a possibility to add 
\bear
I_2 &=&
C_0 \int_{x_4=0} (\wCF + \wCF^\dagger -2\px{4}\wVF)^2 d^4\th\, d^4 x
\nn\\ &&
-\frac{C_1}{4} \int_{x_4=0} \left\lbrack
\int \cW_\a \cW^\a d^2\th
+\int \bcW^\ad \bcW_\ad d^2\bth
\right\rbrack.
\nn
\eear
to the action. Here $C_0$ and $C_1$ are unknown coefficient.
Such a term will not affect the boundary conditions (\ref{Xbc}).

Similarly, we add a chiral superfield $\wCY$ at the other end,
 $x_4 = \pi R$.
The $U(1)$ anomaly cancellation can be satisfied if we 
require that $\wCY$ is also a triplet
and has charge $-1$ under the 5D $U(1)$ gauge group.
We assume that it is in the representation $3$ of another global
$SU(3)$. $\wCY$ satisfies the boundary condition
$$
\frac{1}{2}\chi^2|_{x_4 = \pi R} = |Y|^2,\qquad
Y\equiv \wCY|_{\th=\bth=0}
$$

\subsection{Relation between the linear and nonlinear $\sigma$-models}
% ------------------------------------------------------------------------
We can now solve the equations of motion of the nonlinear
$\sigma$-model at energies $E\ll \frac{1}{R}$.
If we vary the action with respect to $A_4$ and integrate
by parts, keeping boundary terms, so as to extract the coefficient
of $\delta A_4(0)$ we find the equation
$$
0 = \chi^2 A_\u + {\mbox{Im}}(X_i^* D_\u X_i)
+\chi F_{4\u} 
+C_0\qx{4}F_{\u 4}
+C_1\qx{\v}F_{\u\v}.
$$
The first term comes from $|D_\u X|^2$ and the boundary condition
(\ref{Xbc}).
To look for the zero modes we assume that $X_i$,
$A_4$ and $A_\u$ are independent of $x_0,\dots, x_3$.
We find the boundary condition at $x_4 = 0$:
$$
0 = \chi_0^2 A_\u + \chi_0\px{4}A_\u -C_0 \qx{4}\px{4}A_\u.
$$
There is a similar equation at $x_4 = \pi R$.
The equation of motion in the bulk is
$\px{4}\qx{4}A_\u = 0$ and for $\chi_0\neq 0$ there is
no zero mode because any nonzero solution that is independent of
$x_0,\dots,x_3$ cannot satisfy both boundary conditions.

Now we allow the fields to vary as a function of $x_0\dots,x_3$.
For low-energy modes, we can neglect $\qx{\v}F_{\v\u}$
and $\qx{4}\px{4}A_\u$ and $\chi_0\qx{4}A_\u$  compared
to $\chi_0^2 A_\u$. We therefore get
$$
A_\u|_{x_4 = 0} \approx -\frac{1}{\chi_0^2}{\mbox{Im}}(X_i^* D_\u X_i)
+\frac{1}{\chi_0}\px{\u}A_4
$$
and a similar equation at the other end.
Together we find the solution:
$$
A_\u(x_4) \approx
-\frac{\pi R - x_4}{\pi R\chi_0^2}{\mbox{Im}}(X_i^* D_\u X_i)
-\frac{x_4}{\pi R\chi_0^2}{\mbox{Im}}(Y_i^* D_\u Y_i)
+\frac{1}{\chi_0}\px{\u}A_4
$$
Assuming that $R\chi_0\gg 1$ our assumptions about 
neglecting $\chi_0\px{4}A_\u$ are correct.
We also assume that $A_4$ is independent of $x_4$.

\subsection{The moduli space}\label{clmodsp}
% ------------------------------------------------------------------------
The low energy effective action in 4D is a nonlinear $\sigma$-model
with a 5 (complex) dimensional target space. The target space
can be described as a line bundle over $P^2\times P^2$, the same as
the one discussed in subsection (\ref{subsec:modsp}).
The homogeneous coordinates on the first $P^2$ are given by
the fields $X_i/\|X\|$ ($i=1\dots 3$) on the boundary at $x_4 = 0$
and the homogeneous coordinates on the second $P^2$ are
given by $Y_i/\|Y\|$. The coordinates on the fiber of the line bundle
is  the $\th=\bth=0$ component of the superfield
$e^{\int_0^{\pi R}\wCF(x_4) dx_4}$.
We will denote it by $z$.
The symmetry group $SU(3)\times SU(3)$ acts non-trivially on the moduli
space and the orbit of any point is an $S^1$ bundle over $P^2\times P^2$.

Finally, let us discuss the metric on the moduli space.
For $R\chi_0\gg 1$, the size of each $P^2$ is given by
$$
\|X\| = \|Y\| = \frac{1}{\sqrt{2}}\chi_0 = 
\frac{1}{\sqrt{2}\pi R}\log |z|.
$$
The metric on the fiber is
$$
\frac{\chi_0}{\pi R}\frac{|dz|^2}{|z|^2}
=
\frac{\log |z|}{\pi^2 R^2 |z|^2}|dz|^2.
$$

\subsection{The regime opposite to $\chi_0\gg\frac{1}{R}$}\label{conjecture}
% ------------------------------------------------------------------------
If the condition $\chi_0\gg \frac{1}{R}$ is not met then $\chi_0$,
being defined as the 5D VEV, is not well defined.
We cannot use the low-energy effective action (\ref{ezle}) but 
we will propose below an alternative
Lagrangian that describes the compactified theory
at energies $E\ll \frac{1}{R}$.
There are several examples of 5D and 6D strongly interacting
theories that after compactification to 4D are described at low energies
by ordinary field theories. The $(2,0)$ theory, for example, is described
by $N=4$ Super-Yang-Mills theory after compactification on $T^2$ and
at energies below the compactification scale.
We can hope that the $E_0$ theory compactified on the segment $S^1/Z_2$
(with the boundary conditions implied by the M-theory construction above)
is also described by a regular field theory at energies below the 
compactification scale. The heterotic string analysis that
was discussed in section (\ref{subsec:spectrum}) provides a clue.

Before tackling the entire spectrum up to scales $E\ll \frac{1}{R}$ let
us consider the lowest end of the spectrum, namely the moduli space.
The starting point is the classical result of section (\ref{clmodsp}).
$N=1$ supersymmetry allows for quantum corrections to both the
K\"ahler  metric as well as the superpotential.
However, we assume that the global $SU(3)\times SU(3)$ remains a good
symmetry (as the embedding into string theory suggests).
Since the $SU(3)\times SU(3)$ 
orbit of a point in the classical 5(complex) dimensional moduli
space found in section (\ref{clmodsp}) is of real codimension $1$
there cannot be any generated superpotential (since there is no nonzero
holomorphic function that is constant on a real codimension $1$ subspace
and zero at infinity). The K\"ahler potential, on the other hand,
can receive quantum corrections.
The $SU(3)\times SU(3)$ symmetry imposes restrictions on the possible
K\"ahler metrics. To see what the restrictions are it is convenient
to parameterize the moduli space by a complex $3\times 3$ matrix, 
$\Phi_1$, of rank $1$. When the variables $\wCX,\wCY$ and $\wCF$ discussed
in section (\ref{LinSM}) are valid, $\Phi_1$ can be taken as the $\th=\bth=0$
component of the superfield 
\be\label{PhDefXY}
R \wCX e^{\int_0^{\pi R}\wCF(x_4) dx_4}\wCY^T.
\ee
The only $SU(3)\times SU(3)$ invariant that can be constructed from
a rank $1$ matrix is $\xi\equiv\tr{\Phi_1^\dagger\Phi_1}$.
Note that in the regime $\chi_0\gg \frac{1}{R}$ we have
$$
\xi\equiv \tr{\Phi_1^\dagger\Phi_1} = R^4 \chi_0^4 e^{2 \pi R \chi_0}.
$$
The $SU(3)\times SU(3)$ invariant K\"ahler metric can only be a function
of the real variable $\xi$. In the regime ($\xi\gg 1$)
the K\"ahler function is proportional to $\chi_0^3$ 
as can be seen from (\ref{ezle}) (and is a complicated
function of $\xi$). $K(\xi)$ receives quantum corrections and we 
wish to extrapolate to the region of small $\xi$.
We will assume that there is a point $\xi=0$ in the moduli space
where the $SU(3)\times SU(3)$ symmetry is restored.
In principle, the point $\xi=0$ might be infinitely far away on the moduli
space but this seems unlikely. 

If such a point where $SU(3)\times SU(3)$ is restored exists we might hope
that it is described at low energies by an ordinary field theory.
We conjecture that this is indeed the case and that for $\xi\ll 1$
the dynamics at energies $E\sim \frac{\xi^\frac{1}{2}}{R}\ll \frac{1}{R}$ 
is reproduced by the following Lagrangian.
First, we should elevate the field $\Phi_1$ 
defined in (\ref{PhDefXY})
to a generic $3\times 3$ matrix, $\Phi$,
without any restriction on the rank.
The Lagrangian is then
\be\label{conj}
\int \tr{\Phi^\dagger\Phi} d^4\th
+\lambda\int \det\Phi d^2\th
+\lambda\int \det\Phi^\dagger d^2\bth
\ee
Here $\lambda$ is an unknown real constant.

We have seen in section (\ref{subsec:Fterm}) that the potential of the
scalar component of $\Phi$ in
(\ref{conj}) has a minimum when the $3\times 3$ matrix has rank $1$
(or $0$).
The moduli space of $3\times 3$ matrices of rank $1$ can be parameterized
as in (\ref{uvdag}) and using (\ref{PhDefXY}) we see that the massless
spectrum is reproduced correctly.
In other words, at energies 
$E\ll \left\| \sqrt{\langle \tr{\Phi^\dagger\Phi}\rangle}\right\|$
the dynamics that is described
by (\ref{conj}) and the dynamics that is described in (\ref{LinSM})
coincide.

The global symmetry $SU(3)\times SU(3)$ acts on $\Phi$
as $\Phi\rightarrow\Lambda_1\Phi\Lambda_2$ with 
$(\Lambda_1,\Lambda_2)\in SU(3)\times SU(3)$.
Up to an $SU(3)\times SU(3)$ transformation we can choose the VEV
$\langle\Phi\rangle$ to be of the form
$$
\langle\Phi\rangle = 
\left(\begin{array}{ccc}
\phi_0 & 0 & 0 \\ 0 & 0 & 0 \\ 0 & 0 & 0 \\
\end{array}\right).
$$
It breaks $SU(3)\times SU(3)$ down to $SU(2)\times SU(2)\times U(1)$,
as discussed in (\ref{subsec:Fterm}).
The 9 complex fields that comprise $\Phi$ decompose under
$SU(2)\times SU(2)\times U(1)$ as
$$
(1,1)_0 + (2,2)_0 + (1,2)_3 + (2,1)_{-3}.
$$
Here the subindex denotes the $U(1)$ charge.
The 4 fields in the representations $(2,2)_0$ are
massive with masses of order $\phi_0$.
Note that if a particle with the quantum numbers $(2,2)_0$ existed
it could decay into the massless particles 
with quantum numbers $(1,2)_3$ and $(2,1)_{-3}$.
Moreover, we expect $\lambda$ to be a parameter of order $1$ and
therefore (\ref{conj}) is strongly coupled.

The conjecture (\ref{conj}) is motivated by the string theory effective 
action. The string theory derivation assumes, of course, that
$\lam_s^{\frac{1}{3}}= M_p R \ll 1$
 but we conjecture that the form above remains valid even
when $M_p R\gg 1$. The reason is that we expect the 4D low energy
effective action to be scale invariant at energies well below
$\frac{1}{R}$.  The F-term in the expression (\ref{conj}) receives
no quantum corrections and therefore the dimension of $\Phi$
can receive no quantum corrections. 
The K\"ahler term can receive quantum correction. By dimensional 
analysis it should be of the form $\Phi^\dagger\Phi f(R\Phi,R\Phi^\dagger)$
where $f$ is some function of the dimensionless quantities 
$R\Phi$ and $R\Phi^\dagger$. Assuming that  $f(0,0) \neq 0$ we
can normalize $\Phi$ such that $f(0,0)=1$ and
for small $R\Phi$ we can approximate $f$ to be a constant.

Note that $\Phi$ has 9 components but when $\Phi$ gets a nonzero
VEV that is a matrix of rank $1$ only 5 components of $\Phi$
remain massless (since a rank $1$ matrix can be put in the form
(\ref{uvdag}) with the equivalence relations (\ref{uvch})).
Therefore, 4 components of $\Phi$ are massive at a generic
point of the moduli space. The masses of these components are of
the order of $\chi_0$ and when $\chi_0\sim\frac{1}{R}$ the masses
are of the order of the compactification scale. Since we are 
neglecting any modes with masses of the order of $\frac{1}{R}$,
the form (\ref{conj}) does not contain any information
in addition to the moduli space for $\chi_0\sim\frac{1}{R}$.

\section{ Anomalies }
% ========================================================================
In this section we discuss the cancellation of anomalies of local symmetries
in the 4D field theories associated with the  
 $Z_3$ orbifold models. We consider here the case of  
$[E_6\times SU(3)]\times [E_6\times SU(3)]$ gauge symmetry, and the rest of 
the $Z_3$ models in the appendix.
 We analyze the anomalies
  of the low energy effective action of 
  the heterotic string theory 
compactified on compact ${T^6\over Z_3}$,  on non-compact ${C^3 \over Z_3}$,
for  the HW dual model and for the case where the singularity is blown up.

\subsection{The ${T^6\over Z_3}$ orbifold model}
% ------------------------------------------------------------------------
At  each fixed point anomalies can potentially  occur only in the    
$SU(3)\times SU(3)$ subgroup of the local symmetry group. Recall
that the twisted states
are in the  representation $(\bar 3,\bar 3)$ 
so the contribution to the anomaly of  each  $SU(3)$ is that of
3 anti-fundamental representations.
The  charged untwisted massless matter transform in the 
$3(3,27,1,1)\oplus 3(1,1,3,27)$ representation of the full symmetry group.
Thus, the contribution 
 to the anomaly of each $SU(3)$ symmetry group   
is that   of  81 fundamentals  
divided evenly  between all the fixed points, namely divided by 27-
  the number of fixed points.
So altogether the contribution of the 
untwisted matter is that of three  fundamentals.
 Hence there is an exact cancellation between the twisted and untwisted 
states and    
both $SU(3)$ gauge symmetries  are anomaly free.

\subsection{The ${C^3\over Z_3}$ orbifold model} 
% ------------------------------------------------------------------------
Consider now  the heterotic  compactification  on  non-compact 
${R^6\over Z_N}$  Calabi Yau orbifolds.
In such a case, it may seem  that one faces a problem  with the cancellation 
of local anomalies.
Again we consider  the $Z_3$ model with the  unbroken gauge group   
$[E_6\times SU(3)]\times [E_6\times SU(3)]$. 

The content of the massless 
spectrum is the same as  that of the compact case. 
The main difference is  that unlike the 27 fixed points of the compact case
here  
there is only one single singular point, namely,  the origin.  
Thus it seems that there is not reason to divide the contribution of
the untwisted sector by 27 and hence it looks as if the anomaly associated
with the twisted matter cannot balance that of the untwisted  sector.

It turns out that the division that one invokes in the compact case due to
the multiplicity of fixed points, should be implemented also in the 
non-compact case.
As was shown by Gimon and Johnson\cite{GJ},  when performing 
the corresponding one loop stringy  
computation  in the non-compact case,
there is a zero mode of the $Z_3$ projection 
that one has to take into account. 
The trace of the  twist operator $\alpha=e^{2\pi ik\over N}$ that operates on
the complex coordinates $z\rt e^{2\pi ik\over N} z$ for each $T^2$ takes the 
form
$$
Tr[e^{2\pi ik\over N}]=\int dzd\bar z \langle z,\bar z|\alpha 
|z,\bar z\rangle
={1\over {4\sin^2{\pi k\over N}}}
$$  
where one uses the basis with  
$
\langle z|z'\rangle = {1\over V_{T^2}}\delta(z-z')$.
For the  $T^6$  and for a $Z_3$ orbifold  we thus get the factor of  

${1\over (4\sin^2{\pi k\over 3})^3}=\frac{1}{27}$
 which we have to multiply the 
contribution of the untwisted sector.
Therefore, like for the compact case, there is an exact cancellation between
the contributions of the twisted and untwisted sectors to the anomaly. 

\subsection{ The HW dual with blown up fixed point}
% ------------------------------------------------------------------------
In the HW dual of both the compact and non compact orbifold models one has
to cancel the  4D anomalies locally at each point along the $S^1\over Z_2$
interval and in particular at the two ends of it.

We will discuss now the anomaly cancellation for the blown up $R^6/Z_3$ case
and at the end of this subsection we comment on the case where the  fixed points  are not blown up.

Let us first see how the anomalies are canceled when we are at
a generic point of the moduli space where the symmetry is
broken down to $SU(2)_L\times SU(2)_R\times U(1)_V$.
Let us also denote the 5d $U(1)$ gauge group as $U(1)_B$.

The fields $\wCX$ have the following $(SU(2)_L, SU(2)_R)_{(U(1)_V,U(1)_B)}$
quantum numbers:
$$
\wCX: (2,1)_{(1,1)} + (1,1)_{(-2,1)},\qquad
\wCY: (1,2)_{(-1,-1)} + (1,1)_{(2,-1)},
$$
%As usual, the $SU(2)$ quantum numbers denote representations and
%the $U(1)$ quantum numbers denote charges.
We can take the contribution of the untwisted fields to the anomaly
to as that of 3 superfields each with quantum numbers
$$
(2,1)_{(-1,0)} + 
(1,1)_{(2,0)} + 
(1,2)_{(1,0)} +
(1,1)_{(-2,0)}.
$$
When $SU(3)\times SU(3)$ is broken down to $SU(2)_L\times SU(2)_R\times U(1)_V$
the VEVs of $\wCX$ and $\wCY$ are not invariant under $U(1)_V$.
We can rectify this with a compensating $U(1)_B$ gauge transformation.
In other words, we define a $U(1)_C$ such that the charges satisfy
$Q_C\equiv Q_V + 2 Q_B$. The $(SU(2)_L,SU(2)_R)_{U(1)_C}$ quantum numbers
are now:
$$
\wCX: (2,1)_{3} + (1,1)_{0},\qquad
\wCY: (1,2)_{-3} + (1,1)_{0}.
$$
We can now take the VEVs to be invariant under $U(1)_C$.
We can now check the local cancellation of the anomaly.
The fields $\wCX$ together with the untwisted fields  on the $x_4 = 0$
end with quantum numbers $(2,1)_{-1}+ (1,1)_{2}$ cancel the 
$U(1)_C\cdot SU(2)_L^2$ anomaly. However the $U(1)_C^3$ anomaly is not
canceled. 
We get a net $54 F^3$ from $\wCX$ and $18 F^3$ from
the ``untwisted sector'' adding up to $72 F^3$. Here $F$ stands for
a $U(1)_C$ gauge field.
The bulk 5D Chern-Simons term also contributes to the anomaly
and the contribution is $-24 F^3$. Altogether we get $48 F^3$.
The fields at the other end, $x_4 = \pi R$ contribute $-48 F^3$.
Therefore, locally in 5D the $U(1)_C^3$ anomaly does not cancel.

But in fact the $U(1)_C^3$ anomaly is not required to
cancel locally in 5D.
Only the $SU(2)^2\times U(1)_C$ anomalies have to cancel locally.
The $U(1)_C$ gauge transformation is the same with opposite signs on
both ends. In other words, we are not allowed to make a different
$U(1)$ transformation at $x_4=0$ and at $x_4=\pi R$ without changing
the vacuum.

For the HW scenario without blowing up the fixed points, 
and with $R\gg 1/M_p$ we can consider the  
4D world volume theory on each  end of the world brane
neglecting the influence of the physics at the other end. 
Consider the theory  on the left end. This theory has a
gauge symmetry with the gauge group  $[E_6\times SU(3)]_L$.
The untwisted sector still contributes ${1\over 27}$ of the massless $[3(3,27)]$
representation.  Anomaly cancellation, thus, requires three multiplets of 
$3$ of $SU(3)_L$.
  Since we do not have a handle on the structure of the theory 
without a blow up, we can only conjecture on how such  a cancellation 
may occur. One possibility is that the  $\Phi$ 
  field which transforms in  $(\bar 3,\bar 3)$ of the global  
$SU_L(3)\times SU_R(3)$ symmetry group in the bulk
couples at the 4D left theory 
to the $SU_L(3)$ gauge fields
 and the $SU_R(3)$ un-gauged degrees of freedom are flavor degrees
of freedom.
These field cancels the $SU(3)_L$ anomaly of the untwisted fields.
A similar mechanism might take
 place on the right end of the world 4D theory.
Of course, it might be that the field $\Phi$ is only a low-energy
effective description of more fundamental degrees of freedom
and in the regime $R\rightarrow \infty$ the anomaly cancellation
mechanism is different.

\section{Summary and discussion}
%  ========================================================================
In this paper we have addressed  the strong coupling dynamics of the 
$T^6/Z_3$ ( $C^3/Z_3$)  heterotic orbifolds.
The main tool used has been the duality  with the strongly coupled heterotic string theory and the weakly coupled 
Horava Witten M-theory compactified on $(S^1/ Z_2)\times (T^6/ Z_3)$.

The motivation for this study has been three folded: to shed additional light on M-theory 
from the heterotic string, to explore the domain of large string coupling using the Horava-Witten picture 
and to better understand the effective action of the recently popular scenarios of
5D bulk space-time  with two end-of-the-world branes.
 We concentrated mainly on the $Z_3$ orbifold 
that breaks the $E_8\times E_8$ gauge group down to
$SU(3)\times E_6\times SU(3)\times E_6$. 

 We showed that the moduli space  of the   $(C^3/ Z_3)$ heterotic orbifold  is
a blow-down at the zero section of a line bundle that
is the product of the two universal line bundles,
$\Lbd_1\otimes\Lbd_2$,
over $P^2\times P^2$.
The properties of the moduli space were reproduced 
 from the supergravity description in the large blow-up limit.
The gauge instantons of  the symmetric  model as well as of  other
 $T^6/Z_3$ orbifolds were analyzed. 
In the context of the $E_0$ theory on a segment 
we identified   two scales of the system, namely, the compactification scale and the scale of the
expectation value of the scalar field.
We wrote  down the
 $N=1$ supersymmetric 5D  $E_0$  theory  in terms of a 4D $N=1$ chiral and vector superfields.
We then compactified  this theory on $(S^1/ Z_2)$ first in the limit of an expectation value which is   
much larger  than the inverse of the compactification scale.  In this regime we reduced the 
11D HW supergravity to that of a 5D theory  in the form of a non-linear sigma model. We then rewrote
it in terms of a linear sigma model and determined
the relations between the linear an non-linear descriptions.  
 Finally we were led  to a conjecture about the
low energy description of the five dimensional 
$E_0$-theory (the CFT that describes
the the singularity region of M-theory on $C^3/Z_3$)
compactified on $S^1/Z_2$.
%({\bf Cobi, perhaps we can omit the following two sentences?})
%We discussed 
%the cancellation of anomalies of local symmetries
%in the framework of the compact as well as the non-compact 
%heterotic orbifold. 
%It is then argued that the HW theory with the blown up singularities,
%due to the breaking of the 
%symmetry there is in fact no issue of anomalies.

The status of the 
heterotic compactifications on singular CY orbifolds stand
in contrast to 
the situation with compactifications on singular $K3$
 orbifolds. There the HW theory is weakly coupled even before the
singularities are blown-up \cite{KSTY,GKSTY}.
Unfortunately,
we still do not possess a fully coherent picture for the analogous 
models with compactification on singular CY orbifolds.

Another open direction is the search for possible viable phenomenological models on the 4D end-of-the-world branes.
For instance one may introduce Wilson loops to get symmetry breaking patterns that are compatible with the standard model
symmetries.

%===================================================================
%\section*{Acknowledgments}
%===================================================================
\acknowledgments
 We would like to thank Eric Gimon, S. Theisen  and E. Witten for 
 useful discussions and comments.
Our discussion in section (\ref{ModSugra}) in particular benefited
from conversations with E. Witten.
 We would especially like to thank Vadim Kaplunovsky
who took part in the first stages of this project. 
The research of JS was supported in part 
by the US-Israel Binational Science
Foundation, by GIF -- the German-Israeli Foundation for Scientific Research,
and by the Israel Science Foundation. 
The work of OJG was supported in part by the Director, Office of Science,
Office of High Energy and Nuclear Physics, of the U.S. Department of
Energy under Contract DE-AC03-76SF00098, and in part by
the NSF under grant PHY-0098840.
\newpage

\def\np#1#2#3{Nucl.Phys.{\bf B{#1}}({#2}) {#3}}

\newpage
\appendix
\setcounter{equation}{0}
\renewcommand{\theequation}{A-\arabic{equation}}

\section{The spectra and anomalies of the  $Z_3$  heterotic models}
%\section{The spectra of the  $Z_3$  heterotic models}
% ========================================================================= 

We will now describe the spectra for the various orbifolds
of the $E_8\times E_8$ heterotic string theory that were discussed
in the paper. The details can be found in \cite{PolBook}.
(See also \cite{deBoer:2001nw} for a new discussion on orbifolds.)

{}From the target space point of view the different orbifolds are
characterized by the different embeddings of $Z_3$ into the gauge groups
$E_8\times E_8$.
{}From the worldsheet point of view,
the different
orbifoldings are  characterized by two vectors $\vec r = (r_1, r_2, r_3)$
and the shift vector $\vec s= ( s_1,...s_{16})$.
The conditions that these two vectors have to satisfy are
\be 
\sum_i r_i = \sum_i s_i= 0\  mod 2\qquad
\sum_i r_i^2 - \sum_i s_i^2= 0\  mod 6
\ee
In particular for  $ r_i = (1,1,-2)$ the condition takes the form 
 $\sum_i s_i^2= 0 mod6$ 
The nontrivial solutions of these conditions and the corresponding 
gauge groups are 
\bear
(0^8; 0^8) &\rightarrow& [E_8]\times [E_8]\cr
(1,1-2,0^5; 0^8) &\rightarrow& [E_8]\times [E_6\times SU(3)]\cr
(1,1-2,0^5;1,1,-2 0^5) &\rightarrow& E_6\times SU(3)\times[E_6\times SU(3)]\cr
(1,1,0^6;-2 0^7) &\rightarrow&[E_7\times U(1)]\times[SO(14)\times U(1)]\cr
 (1,1,1,1,2,0^3;2 0^7) &\rightarrow&[ SU(9)]\times[SO(14)\times U(1)]
\eear 

The corresponding  spectra of the  models  with broken $E_8$ symmetries are
\bear
 U: 3(1;27, 3)  + 9 moduli&\qquad & 
T: 27[(1;27,1)+ 3(1;1,\bar 3)] \cr
U:  3[(27,3;1,1) + (1,1;27, 3)] + 9 moduli&\qquad &
T:  27(1,\bar 3;1,\bar 3)\cr
 U: 3[(56;1)+ 2(1;1) +(1;64)+ (1;14)]  + 9 moduli&\qquad &
T:  27[(1;14)+ 2(1;1)]\cr
U:  3[(84;1) +(1;64)+ (1;14)]  + 9 moduli&\qquad &
T:  27[(1;\bar 9) ]
\eear
where $U$ adn $T$ stand for the untwisted and twisted sectors respectively. 

Anomalies of the 4D field theory were discussed in section  5 for the  
$E_6\times SU(3)\times[E_6\times SU(3)]$. The same anomaly cancellation mechanism applies also for the right symmetry group $[E_6\times SU(3)]$ of the second
model of the above list. Since $E_7$ and $SO(14)$ are anomaly free groups,
we have to discuss only the anomalies of the $SU(9)$ of the last model
and  of the $U(1)$s of the fourth and fifth
models.
The contribution of the matter in the $84$ representation of $SU(9)$ is the same as that of  9 fundamentals. Since it is part of the untwisted sector it has
to be divided by 27 so that the net contribution is that of a fundamental,
since they come with a multiplicity of three. 
The contribution of the twisted sector is of one anti-fundamental thus one has a
full anomaly cancellation.

  \end{document}